\documentclass[aps,prb,showpacs,twocolumn]{revtex4}
\usepackage{graphics,bm}
\usepackage{amssymb}
\usepackage{epsfig}
\usepackage{epsf}
\usepackage[usenames]{color}
\usepackage{upgreek}
\usepackage{amsmath}
\usepackage{xcolor}
\usepackage{hyperref}

\begin{document}

\title{Magnetic field dependence of the magnon spin diffusion length in the magnetic insulator yttrium iron garnet}
\author{L.J. Cornelissen}
\email{l.j.cornelissen@rug.nl}
\affiliation{
    Physics of Nanodevices, Zernike Institute for Advanced Materials,
    University of Groningen,
    Nijenborgh 4,
    9747 AG Groningen,
    The Netherlands
}

\author{B.J. van Wees}
\affiliation{
    Physics of Nanodevices, Zernike Institute for Advanced Materials,
    University of Groningen,
    Nijenborgh 4,
    9747 AG Groningen,
    The Netherlands
}

\begin{abstract}
We investigated the effect of an external magnetic field on the diffusive spin transport by magnons in the magnetic insulator Y$_3$Fe$_5$O$_{12}$ (YIG), using a non-local magnon transport measurement geometry. We observed a decrease in magnon spin diffusion length $\lambda_m$ for increasing field strengths, where $\lambda_m$ is reduced from 9.6$\pm1.2$ $\upmu$m at 10 mT to 4.2$\pm0.6$ $\upmu$m at 3.5 T at room temperature. In addition, we find that there must be at least one additional transport parameter that depends on the external magnetic field. Our results do not allow us to unambiguously determine whether this is the magnon equilibrium density or the magnon diffusion constant. These results are significant for experiments in the more conventional longitudinal spin Seebeck geometry, since the magnon spin diffusion length sets the length scale for the spin Seebeck effect as well and is relevant for its understanding.
\end{abstract}


\maketitle
The magnetic insulator yttrium iron garnet (YIG) provides an ideal platform for the study of spin waves\cite{Serga2010}, due to its low magnetic damping\cite{Kruglyak2010} and the fact that no electronic currents can flow in this material. It has been shown that spin waves in the GHz regime can be transported through YIG waveguides over large distances.\cite{Chumak2012,Chumak2015} Recently, research efforts are also directed to the high-frequency part of the spin wave spectrum, studying the diffusive transport of quantized spin waves (magnons). This has been largely motivated by the observation of the spin Seebeck effect (SSE) in YIG by Uchida \emph{et al.} [\onlinecite{Uchida2010}], in which a magnon current is generated by applying a temperature gradient over the magnetic insulator. This temperature gradient results in excitation and diffusion of thermal magnons, which can result in thermal spin pumping when the magnetic insulator is coupled to a normal metal layer.\cite{Xiao2010} Very recently it has been shown that these thermal magnons can also be excited electrically, and can transport spin through YIG. Their transport can be described diffusively, characterized by the magnon spin diffusion length $\lambda_m$, the length scale over which the magnon spin current decays exponentially.\cite{Cornelissen2015}

The SSE in YIG has been studied extensively, both theoretically\cite{Adachi2013,PhysRevB.88.064408,PhysRevB.89.014416,Xiao2010} and experimentally.\cite{Giles2015,Kehlberger2015,Jin2015,Kikkawa2015,Kirihara2012,PhysRevB.88.094410,Vlietstra2014,Schreier2013} Recent experiments show that the voltage resulting from the SSE is reduced upon increasing the external magnetic field.\cite{Jin2015,Kikkawa2015,Ritzmann2015} A mechanism in which low-frequency magnons contribute more to the SSE than high-frequency ones has been proposed to explain these results. The magnetic field will open a Zeeman gap in the magnon density of states, thus 'freezing out' the low-frequency magnons with energies below the gap. This could then cause the reduction in SSE signal.

In this paper we investigate the effect of the applied magnetic field on the diffusive transport of magnon spins. We employ a non-local measurement geometry in which we measure the magnon spin signal as a function of distance, which allows us to directly extract the magnon spin diffusion length for various magnetic field strengths. The main advantage of this method is that the locations of both magnon injection and detection are well determined, due to the localized magnon injection and detection resulting from the exchange interaction between a spin accumulation in the platinum injector/detector and magnons in the YIG. This means that the distance over which the magnon spin current diffuses is known precisely. Our results clearly indicate that the magnon spin diffusion length decreases for increasing magnetic field strength, causing a strong reduction of the magnon spin signal.

\begin{figure}
	\includegraphics[width=8.5cm]{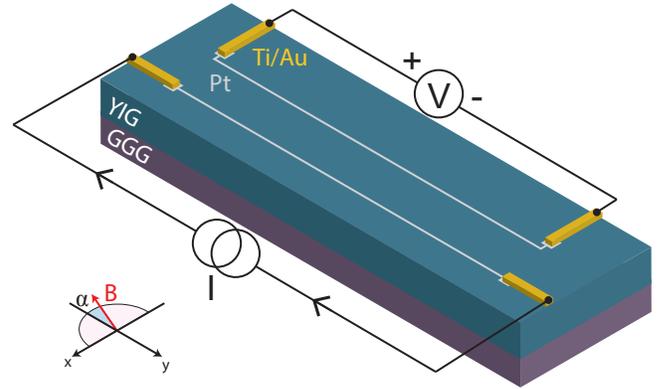}
	\caption{(Color online) Schematic of our typical device geometry. The platinum injector and detector strips are contacted by Ti/Au leads, and current ($I$) and voltage ($V$) connections are indicated. The magnetic field is rotated in the xy-plane (the plane of the sample surface), making an angle $\upalpha$ with the negative y-axis.
	}
	\label{fig:schematic}
\end{figure}

\begin{figure*}
	\includegraphics[width=17.5cm]{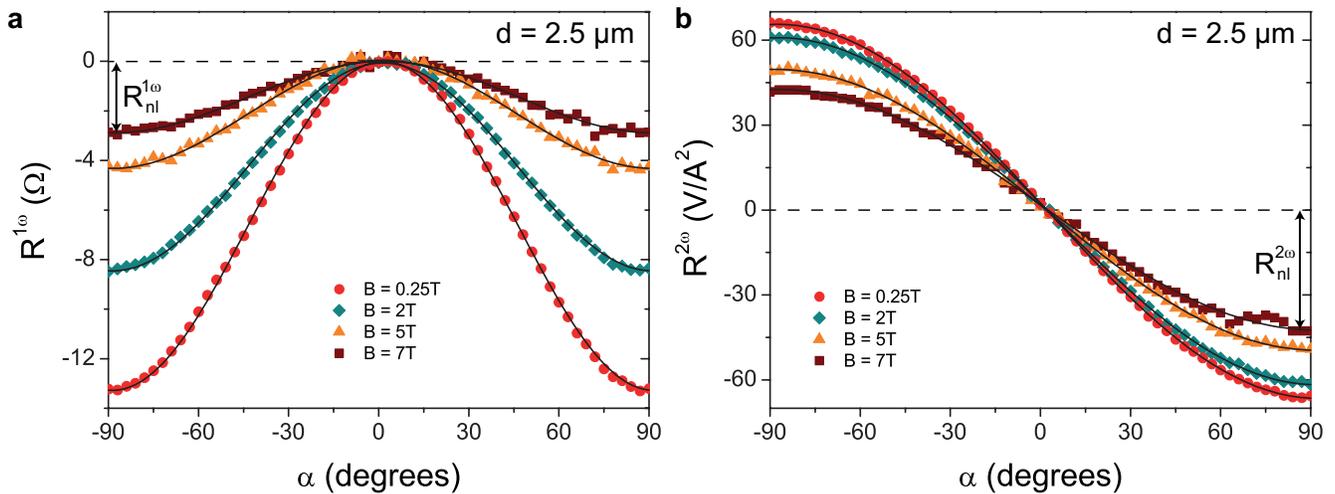}
	\caption{(Color online) Non-local signal as a function of angle $\upalpha$, for an injector-detector separation distance $d=2.5$ $\upmu$m and for various external magnetic field strengths. (a) First harmonic signal. The solid lines are $\sin^2\alpha$ fits through the data. (b) Second harmonic signal. The solid lines are $\sin \alpha$ fits through the data. The decrease in signal amplitude for increasing magnetic field strength is clearly visible for both first and second harmonic signals. The amplitudes of the non-local signals, $R_{\rm nl}^{1\omega}$ and $R_{\rm nl}^{2\omega}$, are indicated in figure (a) and (b) respectively, for $B=7$ T. These measurements were performed using $I_{\rm rms}=110$ $\upmu$A at a lock-in frequency of $f=10.447$ Hz.
	}
	\label{fig:angle_scan}
\end{figure*}
The measurement geometry is shown schematically in Fig.~\ref{fig:schematic} and is equivalent to the non-local geometry we developed in Ref.~[\onlinecite{Cornelissen2015}]. The platinum injector and detector are placed a distance $d$ apart. We measured two series of samples, series A and series B, tailored to perform measurements in the short ($d\sim0.2-5$ $\upmu$m) and long ($d\sim2.5-30$ $\upmu$m) separation distance regime, respectively. Our samples consist of a (111) single crystal Y$_3$Fe$_5$O$_{12}$ film with a thickness of 200 nm (series A) or 210 nm (series B) grown on a 500 $\upmu$m thick (111) Gd$_3$Ga$_5$O$_{12}$ substrate by liquid-phase epitaxy. The YIG samples were provided by the Universit\'e de Bretagne Occidentale in Brest, France (series A) and obtained commercially from Matesy GmbH (series B). We define the platinum injector and detector strips on top of the YIG film using three steps of electron beam lithography. The first step results in a pattern of Ti/Au markers, used to align the following steps. In the second step, we define the platinum injector and detector, which are deposited by DC sputtering in an Ar$^+$ plasma. The platinum thickness is approximately $13.5$ and $7$ nm, for series A and B respectively. In the final step we define Ti/Au (5/75 nm) contacts and bonding pads using electron beam evaporation. Prior to the titanium evaporation, argon ion milling was performed to remove polymer residues from the platinum strips. The platinum injector and detector dimensions are $w^A=100-150$ nm, $w^B=300$ nm, $L^A=12.5$ $\upmu$m and $L^B=100$ $\upmu$m, where $w$ and $L$ denote strip width and length, respectively.

We perform a non-local measurement by applying a current $I$ (typically $I_{\rm rms}=200$ $\upmu$A) to the injector. Due to the spin Hall effect (SHE), this generates a spin current towards the YIG, resulting in a spin accumulation at the YIG\textbar{}Pt interface. Depending on the orientation of the spin accumulation with respect to the YIG magnetization, magnons will be generated in the YIG. These magnons will diffuse to the detector, where they are absorbed and generate a spin current into the YIG, which by virtue of the inverse spin Hall effect will be converted to a charge voltage $V$ (in an open circuit geometry), as shown in Refs.~[\onlinecite{Cornelissen2015, Goennenwein2015}]. The non-local resistance is now defined as $R^{n\omega}=V/I^n$ and is a measure for the magnitude of the magnon spin current between injector and detector. Using a lock-in detection technique\cite{Vlietstra2014}, we are able to separately detect the first harmonic ($n=1$) and second harmonic ($n=2$) response of the sample to our excitation frequency $\omega=2\pi f$, allowing us to separately probe the physics of magnons that are excited electrically\cite{Cornelissen2015, Goennenwein2015} and thermally\cite{Cornelissen2015, Giles2015}, respectively.

We then rotate the sample in an external field, thereby varying the angle between the YIG magnetization and the spin accumulation in the injector. When $\alpha=0$, the magnetization is parallel to the charge current in the injector, hence perpendicular to the spin accumulation and no magnons are excited or detected. When $\alpha=\pm90$ degrees, spin accumulation and magnetization are collinear and the magnon generation and detection efficiency is maximal. The magnitude of the external field is varied, ranging from 10~mT to 7~T. A typical measurement result (for $d=2.5$ $\upmu$m) is shown in Fig.~\ref{fig:angle_scan}, for both the first harmonic (Fig.~\ref{fig:angle_scan}a) and second harmonic (Fig.~\ref{fig:angle_scan}b) response. 

For electrically excited magnons, both injection efficiency $\eta_{\rm inj}$ and detection efficiency $\eta_{\rm det}$ depend on the angle $\alpha$ as $\eta_{\rm inj}\, , \eta_{\rm det} \propto \sin \alpha$. Since the total signal is then proportional to the product of $\eta_{\rm inj}$ and $\eta_{\rm det}$, this results in a total angular dependence of $R^{1\omega}=R^{1\omega}_{\rm nl} \sin^2 \alpha$, where $R^{1\omega}_{\rm nl}$ is the amplitude of the first harmonic signal\cite{Cornelissen2015} (indicated in Fig.~\ref{fig:angle_scan}a). The second harmonic signal however relies on the magnon spin current generated by the spin Seebeck effect in the YIG, due to the temperature gradient arising from Joule heating in the injector. Since Joule heating is independent of $\alpha$, the only angular dependence for the second harmonic non-local signal comes from the magnon detection efficiency $\eta_{\rm det}$, resulting in $R^{2\omega} = R^{2\omega}_{\rm nl} \sin \alpha$, where $R^{2\omega}_{\rm nl}$ is the amplitude of the second harmonic signal\cite{Cornelissen2015} (indicated in Fig.~\ref{fig:angle_scan}b). From Fig.~\ref{fig:angle_scan} we can clearly see that both the first and second harmonic signals decrease for increasing external field strengths.

\begin{figure}
	\includegraphics[width=8.5cm]{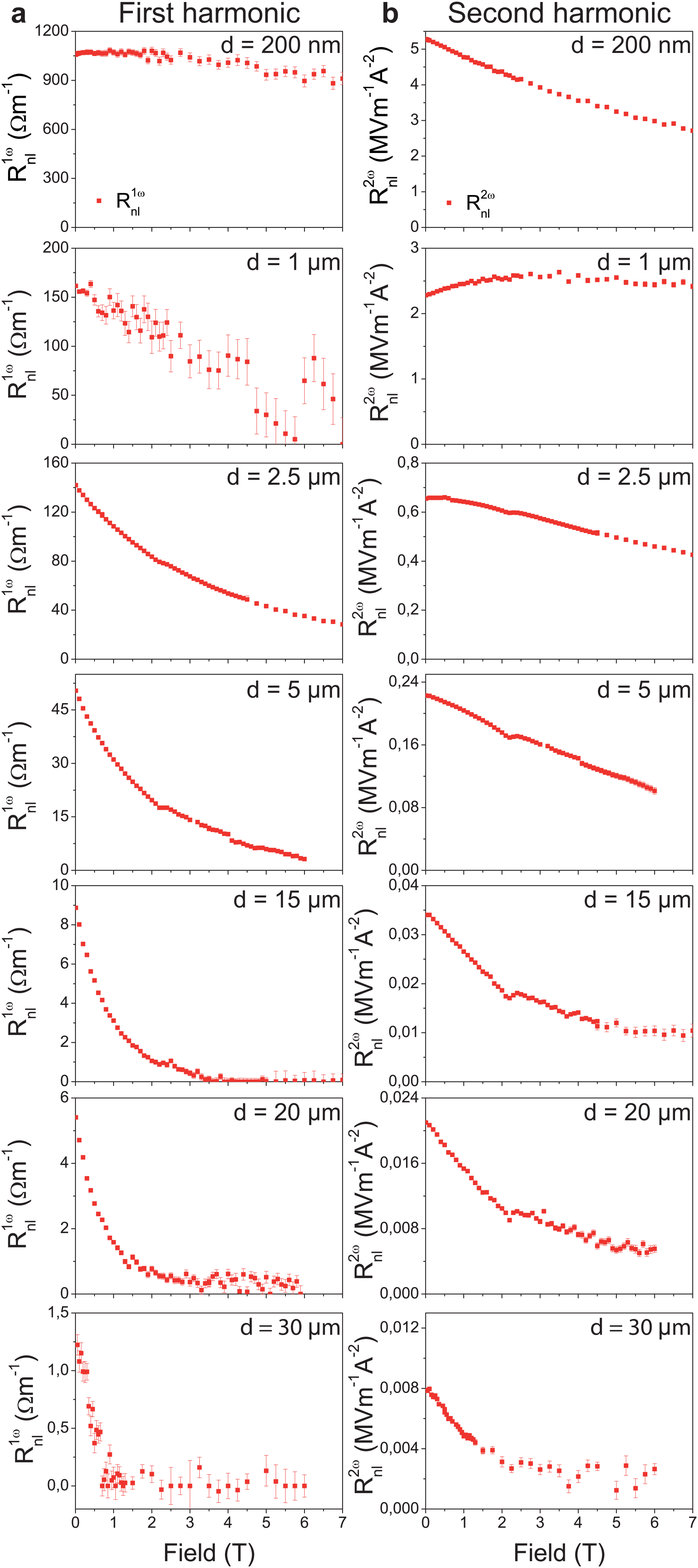}
	\caption{(Color online) Magnitude of the first (a) and second (b) harmonic non-local signals (normalized to device length) as a function of magnetic field, for injector-detector separation distances $d=200$ nm to $d=30$ $\upmu$m. In each plot, the red squares mark the amplitude of the signal, extracted from an angle-dependent measurement as shown in Fig.~\ref{fig:angle_scan}. The errorbars represent the standard error in the fit to extract the amplitude. All measurements were performed at an excitation current of $I_{\rm rms}=200$ $\upmu$A with frequency $f=10.447$ Hz. 
	}
	\label{fig:amplitude_vs_field}
\end{figure}

In order to investigate the dependence of non-local signals on the magnetic field, we performed a series of non-local measurements as a function of field strength for various injector-detector separation distances. The results are shown in Fig.~\ref{fig:amplitude_vs_field}, presenting the data for the first harmonic signal on the left (Fig.~\ref{fig:amplitude_vs_field}a) and the second harmonic signal on the right (Fig.~\ref{fig:amplitude_vs_field}b). The distances that we measured are 200 nm, 1 $\upmu$m, 2.5 $\upmu$m, 5 $\upmu$m, 15 $\upmu$m, 20 $\upmu$m and 30 $\upmu$m. The devices with $d=200$ nm and $d=1$ $\upmu$m are in sample series A, the other distances in series B.

\begin{figure*}
	\includegraphics[width=17.5cm]{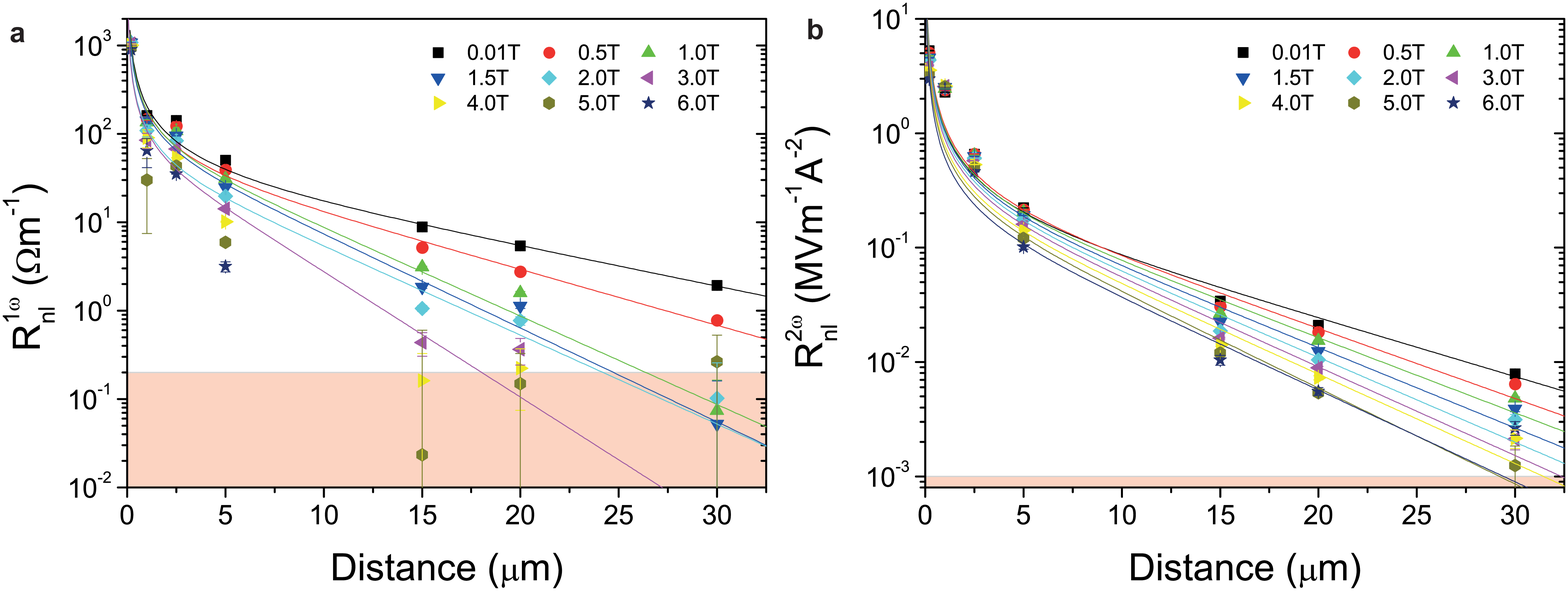}
	\caption{(Color online) Data presented in Fig.~\ref{fig:amplitude_vs_field},  plotted as a function of distance for the first (a) and second (b) harmonic non-local signals (normalized to device length), for magnetic field strengths of $B=0.01$ T to $B=6.0$ T. In each plot, the symbols mark the amplitude of the signal, extracted from an angle-dependent measurement as shown in Fig.~\ref{fig:angle_scan}. The errorbars represent the standard error in the fit to extract the amplitude. The solid lines are fits of Eq.~\ref{eqn:nonlocalsignal} through the data. The shaded regions on the bottom of the graphs indicate the noise floor of the setup for our measurement settings.
	}
	\label{fig:distance_dependence}
\end{figure*}
For both first and second harmonic results, it can be seen that the signal at larger distances is suppressed much more strongly by the external field than at shorter distances. In particular, for the first harmonic response, at $d=30$ $\upmu$m the signal is reduced to $\approx0$ for a field of 1 T, whereas for $d=200$ nm there is virtually no signal reduction up to approximately 1.5 T. For the intermediate distance $d=2.5$ $\upmu$m the signal is suppressed for a field of 1 T, but only by 24\% (compared to the signal at 10 mT). These observations clearly indicate that the mechanism leading to signal suppression must lie in the magnon transport rather than in the generation or detection of magnons: A reduction in $\eta_{\rm inj}$ or $\eta_{\rm det}$ would lead to the same signal suppression at all distances. 

As we derived in Ref.~[\onlinecite{Cornelissen2015}], the non-local resistance as a function of injector-detector separation distance is given by
\begin{equation}
R_{\rm nl}=\frac{C}{\lambda_m}\frac{\exp \left(d/\lambda_m\right)}{1-\exp \left(2d/\lambda_m\right)}\, ,
\label{eqn:nonlocalsignal}
\end{equation}
where $d$ is the distance between injector and detector and $C$ is a distance independent pre-factor that depends for instance on the effective spin mixing conductance of the Pt\textbar{}YIG interface and the magnon diffusion constant $D_m$. Furthermore, $\lambda_m=\sqrt{D_m\tau}$ is the magnon spin diffusion length, where $\tau$ is the magnon spin relaxation time. From Eq.~(\ref{eqn:nonlocalsignal}) it becomes apparent that for $d>\lambda_m$ a slight reduction of $\lambda_m$ can cause a large drop in $R_{\rm nl}$, while as long as $d\ll\lambda_m$ the non-local resistance is (in first order approximation) equal to $-C/(2d)$ and hence the signal will not be influenced by a change in $\lambda_m$. The behaviour observed in the data shown in Fig.~\ref{fig:amplitude_vs_field} can therefore be explained by assuming that $\lambda_m$ is not a constant, yet is reduced under the influence of the external field.

In order to assess the field dependency of $\lambda_m$, we plot the data presented in Fig.~\ref{fig:amplitude_vs_field} as a function of distance, for various magnetic field strengths. This allows us to extract $\lambda_m$ at each field value, by fitting the distance dependent data to Eq.~(\ref{eqn:nonlocalsignal}). The results of this procedure are shown in Fig.~\ref{fig:distance_dependence}a and b for the first and second harmonic signals, respectively. The solid lines are fits through the data to Eq.~(\ref{eqn:nonlocalsignal}), which are performed with weights $w_i\propto1/y_i^2$, where $y_i$ is the amplitude of data point $i$, thus giving more weight to data points at large distances (which have a smaller amplitude but contain more information about $\lambda_m$ compared to the points at short distances). It can be seen from the figure that for both the first and second harmonic, the slope of the fit (in the region $d>5$ $\upmu$m) changes for increasing field strength, indicating a decrease of $\lambda_m$. The shaded regions in the plots represent the noise floor in our measurement setup, which is approximately 4 nV$_{\rm rms}$. We perform a fit of Eq.~(\ref{eqn:nonlocalsignal}) to the data up to $B=3.5$ T since the signal has dropped below the noise floor at that field value for distances $d\geq15$ $\upmu$m, which leaves us with insufficient data points to unambiguously extract $\lambda_m$ for larger field values. 
The same procedure is used for the second harmonic data presented in Fig.~\ref{fig:distance_dependence}b, where in this case we can perform the fits up to $B=6.0$ T due to the larger signal-to-noise ratio for the second harmonic signal.

\begin{figure*}
	\includegraphics[width=17.5cm]{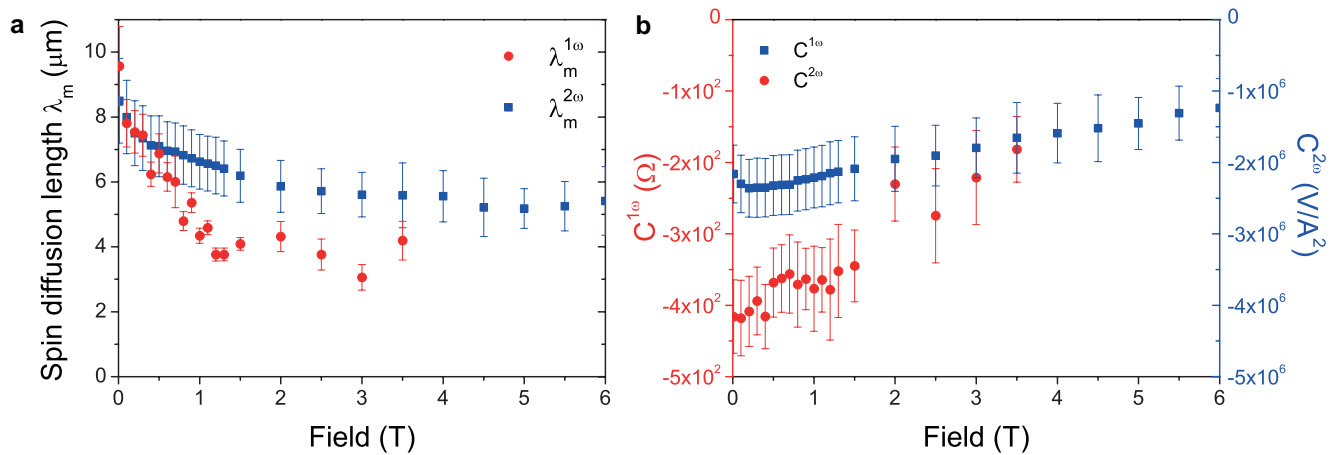}
	\caption{(Color online) (a) Magnon spin diffusion length $\lambda_m$ as a function of external magnetic field, extracted from a fit of Eq.~\ref{eqn:nonlocalsignal} to the distance dependence of the first harmonic (red circles) and second harmonic (blue squares) signals. (b) Pre-factor $C$ as a function of external magnetic field, extracted from the first (red circles, left axis) and second (blue squares, right axis) harmonic signals.
	}
	\label{fig:lambda_vs_field}
\end{figure*}

From the fits shown in Fig.~\ref{fig:distance_dependence} we find the magnon spin diffusion length as a function of field, $\lambda_m(B)$, which we plotted in Fig.~\ref{fig:lambda_vs_field}a. In this figure, both the spin diffusion length extracted from the first harmonic and second harmonic signals ($\lambda^{1\omega}$ and $\lambda^{2\omega}$, respectively) are shown. For fields up to $B=0.7$ T, the spin diffusion lengths extracted from the first and second harmonic signals are equal within the measurement uncertainty. For larger fields however, $\lambda^{1\omega}$ saturates to a smaller value than $\lambda^{2\omega}$. This corresponds to the smaller change in slope of the fits when comparing the distance dependence of the second harmonic signal in Fig.~\ref{fig:distance_dependence}b to that of the first harmonic in Fig.~\ref{fig:distance_dependence}a. This is due to the fact that while the first harmonic signal truly drops to zero for large fields (see Fig.~\ref{fig:amplitude_vs_field}a for $d\geq15$ $\upmu$m), the second harmonic signal saturates at a finite value even for very large field and distance (see Fig.~\ref{fig:amplitude_vs_field}b, for instance $d=30$ $\upmu$m). This finite saturation value might be due to a local heating effect: While at large fields the magnons generated near the injector cannot reach the detector any more due to the short spin diffusion length, a small temperature gradient (resulting from injector Joule heating) could very well still be present near the detector. This temperature gradient would then give rise to a spin Seebeck voltage as it generates magnons locally, meaning that no spin information, but only heat, is transported from injector to detector. This theory could be tested quantitatively by performing detailed finite element modeling of our devices, which would be interesting but is beyond the scope of this current paper.

Another indication that the situation for the thermally excited magnons is more complicated comes from the fact that for $d=1$ $\upmu$m the second harmonic signal slightly increases up to a field of $2.5$ T, rather than immediately decreasing as is observed for all other distances. We do not have an explanation for this behaviour at this moment. 

Finally, from the data for $d=200$ nm it is clear that the non-local signal is reduced for large fields, despite the fact that $d\ll\lambda_m$ which should imply that $R_{\rm nl}$ is independent of $\lambda_m$. Specifically, a reduction of $\lambda_m$ from 9.5 $\upmu$m to 4 $\upmu$m, as shown in Fig.~\ref{fig:lambda_vs_field}a, should result in a signal reduction of only $0.03$\% at $d=200$ nm. The observed signal reduction for this distance is 3\% (from $B=10$ mT to $3.5$ T), which is thus too large to be explained only by a reduction of $\lambda_m$. This can be understood by realizing that the pre-factor $C$ is also reduced under the influence of the magnetic field, as shown in Fig.~\ref{fig:lambda_vs_field}b. Comparing the situation for a magnetic field of 10 mT and 3.5 T, both $\lambda_m^{1\omega}$ and $C^{1\omega}$ are reduced by a factor of 0.44. Since $C^{1\omega}$ depends linearly on $D_m$, we might assume that the reduction of $C^{1\omega}$ can be explained by a reduction of $D_m$ by this same factor. However, since we have that $\lambda_m=\sqrt{D_m\tau}$, this means that $\tau$ also has to decrease as a function of the magnetic field in order to explain the observed change in $\lambda_m$. However, the equilibrium magnon density $n$ also influences $C$, so the effect we observed might also be explained by a reduction of $n$ as proposed in Refs.~[\onlinecite{Jin2015, Kikkawa2015}]. 

Summarizing, we have investigated the influence of an external magnetic field on the diffusive transport of magnon spins in YIG. The most important effect that we found is that the magnon spin diffusion length reduces as a function of field, decreasing from $\lambda_m=9.6\pm1.2$ $\upmu$m at 10 mT to $\lambda_m=4.2\pm0.6$ $\upmu$m at 3.5 T at room temperature. For field values higher than 3.5 T, we cannot extract $\lambda_m$ reliably since the signals at long distances drop below the noise floor for those fields. We also found that for thermally generated magnons, $\lambda_m$ appears to saturate at a higher value than for electrically generated ones. We postulate that this might be due to the presence of a small but finite local contribution to the SSE at the detector, arising from diffusion of the heat generated at the injector. This implies that for large fields, we can no longer rely on the second harmonic signal to extract $\lambda_m$. Furthermore, we showed that the observed signal reduction cannot be explained solely by the suppression of $\lambda_m$, but requires an additional transport parameter to be field-dependent. From the data presented here we cannot identify whether this parameter is the magnon diffusion constant $D_m$ or the equilibrium magnon density $n$. However, it is clear that the observed magnetic field dependence of the magnon spin diffusion length needs to be taken into account in the analysis of the magnetic field dependence of the spin Seebeck effect.

The authors would like to acknowledge H. M. de Roosz, J.G. Holstein, H. Adema and T.J. Schouten for technical assistance, J. Ben Youssef for providing the YIG film used in the fabrication of sample series A and R.A. Duine for discussions. This work is part of the research program of the Foundation for Fundamental Research on Matter (FOM) and supported by NanoLab NL, EU FP7 ICT Grant No. 612759 InSpin and the Zernike Institute for Advanced Materials.
\bibliography{main}
\end{document}